\documentclass[floats,floatfix,showpacs,amssymb,prd,twocolumn,superscriptaddress,nofootinbib,longbibliography,reprint]{revtex4-1}

\usepackage{amssymb,amsmath,verbatim,mathtools,needspace,enumitem,etoolbox,graphicx,physics,microtype,afterpage,bigints,gensymb,tabularx}

\usepackage[dvipsnames, usenames]{xcolor}
\definecolor{linkcolor}{rgb}{0.0,0.3,0.5}
\definecolor{dodgerblue}{HTML}{1E90FF}
\usepackage[unicode, colorlinks=true, linkcolor=linkcolor, citecolor=linkcolor, filecolor=linkcolor,urlcolor=linkcolor, pdfusetitle]{hyperref}
\usepackage[all]{hypcap}
\usepackage[T1]{fontenc}
\usepackage[utf8]{inputenc}
\usepackage{orcidlink}
\usepackage[nolist,nohyperlinks]{acronym}
\usepackage{comment}
\usepackage{multirow}
\usepackage{placeins}
\usepackage{soul}

\newcolumntype{C}[1]{>{\centering\arraybackslash}m{#1}}

\newcommand{\del}{\partial}

\makeatletter
\newcommand*{\balancecolsandclearpage}{\close@column@grid \cleardoublepage \twocolumngrid}
\makeatother

\acrodef{LSC}[LSC]{LIGO Scientific Collaboration}
\acrodef{BH}{black hole}
\acrodef{NS}{neutron star}
\acrodef{PN}{Post-Newtonian}
\acrodef{BBH}{binary black-hole}
\acrodef{MBHB}{massive black-hole binary}
\acrodefplural{MBHB}[MBHBs]{massive black-hole binaries}
\acrodef{BNS}{binary neutron-star}
\acrodef{NSBH}{neutron-star black-hole}
\acrodef{EOB}{effective-one-body}
\acrodef{NR}{numerical relativity}
\acrodef{SNR}{signal-to-noise ratio}
\acrodef{GW}{gravitational-wave}
\acrodef{PSD}{power spectral density}
\acrodef{IMBH}{intermediate-mass black hole}
\acrodef{aLIGO}{Advanced Laser interferometer Gravitational-Wave Observatory}
\acrodef{AZDHP}{aLIGO zero detuned high power density}
\acrodef{GR}{general relativity}
\acrodef{PE}{parameter estimation}
\acrodef{LAL}{LIGO algorithm library}
\acrodef{TPI}{tensor-product interpolant}
\acrodef{SVD}{singular value decomposition}
\acrodef{ODE}{ordinary differential equation}
\acrodef{PDE}{partial differential equation}
\acrodef{ROM}{reduced order model}
\acrodef{QNM}{quasi-normal mode}
\acrodef{LISA}{Laser Interferometer Space Antenna}

\def\apjl{{Astrophys.~J.~Lett.}}
\def\apjs{{Astrophys.~J.~Supp.}}
\def\ao{{Appl.~Opt.}}

\def\aap{{Astron.~Astrophys.}}

\def\baas{{BAAS}}

\def\mnras{{Mon.~Not.~R.~Astron.~Soc.}}

\newcommand{\JHU}{\affiliation{Department of Physics and Astronomy, Johns Hopkins University, 3400 N. Charles Street, Baltimore, Maryland, 21218, USA}}

\begin{document}

\title{Intermediate-mass black hole binary parameter estimation\\with next-generation ground-based detector networks}

\author{Luca Reali$\,$\orcidlink{0000-0002-8143-6767}}
\email{lreali1@jhu.edu}

\author{Roberto Cotesta$\,$\orcidlink{0000-0001-6568-6814}}

\author{Andrea Antonelli$\,$\orcidlink{0000-0001-6536-0822}}

\author{Konstantinos Kritos$\,$\orcidlink{0000-0002-0212-3472}}
\email{kkritos1@jhu.edu}

\author{Vladimir Strokov$\,$\orcidlink{0000-0002-6555-8211}}
\email{vstroko1@jhu.edu}

\author{Emanuele Berti$\,$\orcidlink{0000-0003-0751-5130}}
\email{berti@jhu.edu}

\JHU


\date{\today}

\begin{abstract}
Astrophysical scenarios for the formation and evolution of intermediate-mass black holes (IMBHs) in the mass range $10^2 M_\odot \lesssim M \lesssim 10^6 M_\odot$ remain uncertain, but future ground-based gravitational-wave (GW) interferometers will probe the lower end of the IMBH mass range.  We study the detectability of IMBH binary mergers and the measurability of their parameters with next-generation ground-based detector networks consisting of various combinations of Cosmic Explorer (CE) and Einstein Telescope (ET) interferometers.
We find that, for binaries with component masses $m_{1,2}\sim 1000\,M_\odot$,  an optimal 3-detector network can constrain the masses with errors $\lesssim 0.1\%$ ($\lesssim 1\%$) at $z=0.5$ ($z=2$), and the source redshift can be measured with percent-level accuracy or better at $z\lesssim 2$. The redshift of lighter binaries ($m_{1,2}\lesssim 300\,M_\odot$) can still be measured with $\mathcal{O}(10)\%$ accuracy even at $z=10$. Binaries with $z\lesssim 0.5$ can be localized within $1\,\rm{deg}^2$ for $m_{1,2}\lesssim 1000\,M_\odot$, and within $0.1\,\rm{deg}^2$ for comparable mass systems. The sky localization is good enough that it may be possible to cross-correlate GW searches with galaxy catalogs and to search for electromagnetic counterparts to IMBH mergers. We also point out that the low-frequency sensitivity of the detectors is crucial for IMBH detection and parameter estimation. It will be interesting to use our results in conjunction with population synthesis codes to constrain astrophysical IMBH formation models.

\end{abstract}

\maketitle

\section{Introduction}

Intermediate-mass black holes (IMBHs) span a broad range of masses $10^2 M_\odot \lesssim M \lesssim 10^6  M_\odot$ that fills the gap between stellar-mass black holes (BHs) and supermassive BHs~\cite{Miller:2003sc,Mezcua:2017npy,Greene:2019vlv,Askar:2023pmd}. They are crucial to our understanding of stellar evolution, the seeding of supermassive BHs, and BH-galaxy coevolution~\cite{Volonteri:2010wz,Volonteri:2012tp}. 
Hints of the existence of IMBHs come from ultraluminous X-ray sources~\cite{Greene:2007wy, Kaaret:2017tcn}, pulsar timing~\cite{Prager:2016puh, Abbate:2019qoc}, stellar dynamics~\cite{Gualandris:2009cc, vanderMarel:2009mc, Gualandris:2010qe, Noyola:2010ab, Lutzgendorf:2011dd, Lutzgendorf:2012sz, Lanzoni:2013zfa, Naoz:2019sjx, 2019MNRAS.482.3669G, 2019MNRAS.488.5340B}, and tidal disruption events~\cite{Shen:2013oma, Lin:2016jkb, Chen:2018foj, Fragione:2018lvy, Sakurai:2018lin, Lin:2018dev}.
Despite theoretical models predicting the production of IMBHs in dense star clusters through gravitational runaways~\cite{Miller:2001ez,AtakanGurkan:2003hm,Gultekin:2004pm,PortegiesZwart:2002iks,Mapelli:2016vca,Sedda:2019rfd,Gonzalez:2020xah,DiCarlo:2021att,Rizzuto:2022fdp,Prieto:2024pkt,Purohit:2024zkl,Atallah:2022toy,Kritos:2022non,Kritos:2024upo}, observational evidence for light emission from accreting BHs in the cores of globular clusters remains absent or insignificant~\cite{Pooley:2006aq, Maccarone:2008fw, 2010MNRAS.408.2511M, Lu:2011mya,Miller-Jones:2012trw, Strader:2012nn, Tremou:2018rvq}.

Gravitational-wave (GW) experiments are uniquely positioned to directly probe the IMBH mass range~\cite{Graff:2015bba,Veitch:2015ela}. 
Probing the demographics of IMBHs with a multitude of similar GW signals would provide unique insight into the physics of pair-instability supernovae~\cite{Mehta:2021fgz,Franciolini:2024vis}, the collapse of Population III stars~\cite{Madau:2001sc,Bromm:2001bi,Bromm:2003vv}, the direct collapse of metal-poor gas ~\cite{Loeb:1994wv,Bromm:2002hb,Latif:2013dua,Regan:2017vre}, their growth channel~\cite{Reynolds:2020jwt}, and the properties of star clusters~\cite{IMBH-follow-up}, to name a few.
The LIGO--Virgo--KAGRA collaboration reported at least one event, GW190521, in which the merger remnant mass $M = 142^{+28}_{-16} M_\odot$ is unambiguously within the IMBH range~\cite{LIGOScientific:2020iuh}. The latest GWTC-3 catalog~\cite{KAGRA:2021vkt} also contains a few events with remnant masses that, although smaller, are compatible with the lower end of the IMBH mass range. 
Targeted searches for IMBH signals are ongoing with current GW interferometers~\cite{LIGOScientific:2021tfm}, but injection studies show that these searches are mostly sensitive to binaries with total mass $M \lesssim 200 M_\odot$~\cite{CalderonBustillo:2017skv,Chandra:2020ccy}. This is mainly because current networks are only sensitive at frequencies $\gtrsim 10$\,Hz.
Moreover, based on a search pipeline including waveforms with higher-order harmonics, Ref.~\cite{Wadekar:2024zdq} found hints of a subpopulation of heavy BH mergers with masses up to $\sim300M_\odot$.

Next-generation (XG) ground-based detectors such as the Einstein Telescope (ET)~\cite{Punturo:2010zz} and Cosmic Explorer (CE)~\cite{Evans:2021gyd} will be sensitive at lower frequencies and have an overall higher sensitivity than current interferometers, and as such they will detect BH mergers across cosmic history~\cite{Borhanian:2022czq,Iacovelli:2022bbs,Pieroni:2022bbh}. These detectors are therefore ideal to characterize the lower end of the IMBH mass range. At the higher end, they will be complemented by the Laser Interferometer Space Antenna (LISA)~\cite{LISA:2017pwj,Colpi:2024xhw}. Operating at mHz frequencies, LISA can observe IMBHs~\cite{Amaro-Seoane:2007osp} with significantly higher masses up to redshifts $ z \sim 6$~\cite{Arca-Sedda:2020lso}. There are also proposals to use GWs in the LISA and ground-based detector bands to detect IMBHs via indirect methods~\cite{Kocsis:2011ch, Meiron:2016ipr, Wong:2019hsq, Strokov:2021mkv,Strokov:2023kmo}. Experimental efforts targeting the deciHz frequency band (see e.g.~\cite{10.1063/1.39103, 10.1063/1.39356, Bender:2013nsa, Sato:2017dkf, Coleman:2018ozp, 2019BAAS...51c.453H, AEDGE:2019nxb, Sedda:2021yhn, Branchesi:2023sjl, Ajith:2024mie}) would open a unique GW window into IMBH astrophysics~\cite{Sedda:2019uro,Ajith:2024mie}. 

Typical IMBH merger rates are predicted in the range $\sim 0.01-10$ Gpc$^{-3}$yr$^{-1}$, which suggests from a few to a few thousand detections per year~\cite{Brown:2006pj, Mandel:2007hi, Gair:2010dx,Rasskazov:2019tgb,Fragione:2022avp,Ellis:2023iyb}. Recent results from Pulsar Timing Arrays~\cite{NANOGrav:2023gor} can also be used to constrain the merger rates in the mass range of interest (see e.g.~\cite{2024arXiv240110983I}, which however focuses on LISA rates).
The \emph{detectability} of IMBHs by ground-based GW interferometer networks has been estimated in a few scenarios, including repeated mergers in globular clusters (such as~\cite{Fragione:2017blf,Rasskazov:2019tgb}), nuclear star clusters~\cite{Fragione:2022avp}, and these systems are also interesting because they are ideal ``multiband'' GW sources~\cite{Amaro-Seoane:2009vjl,Cutler:2019krq, Jani:2019ffg,Chen:2022sae}. 
The literature however lacks a systematic assessment of the \emph{measurability} of the properties of IMBHs, such as masses and redshift. Tight constraints on the parameters of these sources would allow to unambiguously place them in the IMBH mass range, as well unlocking the potential for astrophysical population studies. 

Considering older configurations of ET and waveform models, the authors of Ref.~\cite{Huerta:2010tp} showed that it is possible to constrain the parameters of binaries with one IMBH ($100\,M_\odot$ or $500\,M_\odot$) and one stellar-mass object ($1.4\,M_\odot$ or $10\,M_\odot$) with good accuracy When comparing the capabilities of several future ground-based networks from $\rm{A}^\#$ to XG, the authors of Ref.~\cite{Gupta:2023lga} also studied the measurability of source-frame masses for a toy-model population of IMBHs with masses up to $1000\,M_\odot$.
Working with a network of XG detectors, a more recent study performed full Bayesian parameter estimation (PE) for selected high-redshift sources with a total mass of $100-600\,M_\odot$, assessing the measurability of their parameters and the contribution of higher-order harmonics~\cite{Fairhurst:2023beb}.

In this work we perform a more systematic study of the detectability of binaries with at least one IMBH, focusing mostly on the measurability of their parameters, in the context of XG detectors. Since astrophysical merger rates and formation scenarios are very uncertain, we agnostically scan the parameter space by considering grids in BH masses and redshift. We then compute the corresponding signal-to-noise ratios (SNRs) and estimate the inference errors on their parameters using the Fisher information matrix. One of our main goals is to understand which combinations of XG detectors would optimize our ability to infer the properties of IMBHs. With this goal in mind, we compare several different XG detector networks, and we also assess the impact of their low-frequency sensitivity. We present our results by averaging over sky position, polarization, and inclination. 
Being agnostic on the formation scenarios, our parameter estimation results can be used to assess whether XG detectors can observe and measure the properties of binaries with IMBHs resulting from specific astrophysical models.

The rest of the paper is organized as follows. In Sec.~\ref{sec:setup} we summarize our implementation of the Fisher information matrix formalism, along with the different detector networks we consider. In Sec.~\ref{sec:results} we discuss our results for the optimal detector network in our study, and we investigate the degradation in parameter estimation that would result from more pessimistic choices. In Sec.~\ref{sec:conclusions} we draw our conclusions.
In Appendix~\ref{app:singledet} we show examples of how single XG detectors would perform compared to the networks considered in the rest of the study. In Appendix~\ref{app:pe_comparison} we validate our Fisher matrix estimates against Bayesian PE calculations in some specific cases. All data from our Fisher runs and comparisons with full PE can be found at Ref.~\cite{datarelease}. Throughout the paper, we use geometrical units $(G=c=1)$ and we assume the Planck 2018 $\Lambda$CDM cosmological model~\cite{Planck:2018vyg}.

\section{Methods and setup}
\label{sec:setup}

In this section, we describe our methods and setup. In Sec.~\ref{sec:grid} we define the grid we use to explore the binary parameter space. In Sec.~\ref{sec:fisher} we briefly describe our data analysis methods. In Sec.~\ref{sec:networks} we discuss the different detector networks we consider, their low-frequency sensitivity, and the waveform model chosen for this study.

\subsection{Parameter space exploration}
~\label{sec:grid}

Several IMBH formation mechanisms have been proposed in the literature.
Our goal is to assess how well we can constrain the parameters of IMBH binaries with XG detectors, so we do not focus on specific formation scenarios, but we perform an agnostic parameter scan. 

Assuming quasicircular binaries with aligned spins, an IMBH binary system is characterized by an 11-dimensional parameter vector
\begin{equation}
    {\boldsymbol  \theta}=\{m_1, m_2, \chi_{1,z}, \chi_{2,z}, z, \iota,\alpha,\delta, \psi, \varphi_c, t_c\}\,.
\end{equation}
Here, $m_{i}$ are the source-frame masses of the binary components, $\chi_{i,z}$ denotes the magnitudes of the component spins (assumed to be aligned with the binary's orbital angular momentum), $z$ is the redshift, $\iota$ is the inclination angle, $\alpha$, $\delta$ and $\psi$ are the right ascension, declination and polarization angles, $\phi_c$ is the coalescence phase, and $t_c$ is the coalescence time. 

When we assess detectability and measurability for our optimal detector network, we distribute the source-frame masses in log-uniform grids with boundaries $m_1 \in [10^2, 10^4]\,M_\odot$ and $m_2 \in [5, 10^4]\,M_\odot$. Waveform models are calibrated to numerical relativity simulations, which typically do not extend to mass ratios $q = m_1/m_2>10$~\cite{Boyle:2019kee}. For this reason, we only retain configurations with mass ratios $q <100$ when computing SNRs and $q <10$ when estimating parameter uncertainties. We expect the systematic error in current waveform models to yield inaccurate results for larger mass ratios. 

Comparing the results for different detector networks has higher computational cost, so in this case we distribute the total mass $M = m_1 + m_2$ on a log-uniform grid within $[10^2, 10^4]\,M_\odot$ and perform the analysis for two selected values of the mass ratio ($q=1$ and $q=10$). In both cases, we consider redshift values on a uniform grid with boundaries $z \in [0,20]$.

We assume the aligned spin components $\chi_{1,z}, \chi_{2,z}$ to be zero for all the systems we consider. The IMBH spin magnitudes are highly uncertain, as they heavily depend on the formation scenario. However, we are mostly interested in measurements of masses, redshift and sky localization, and we do not expect our results to change dramatically if higher spins were to be considered.

Finally, we distribute both the sky angles $(\alpha,\delta)$ and the orientation angles $(\iota,\psi)$ isotropically. We set the coalescence phase $\varphi_c$ and time $t_c$ to zero for all of our binaries.

\begin{figure}
\centering
\includegraphics[width=\linewidth]{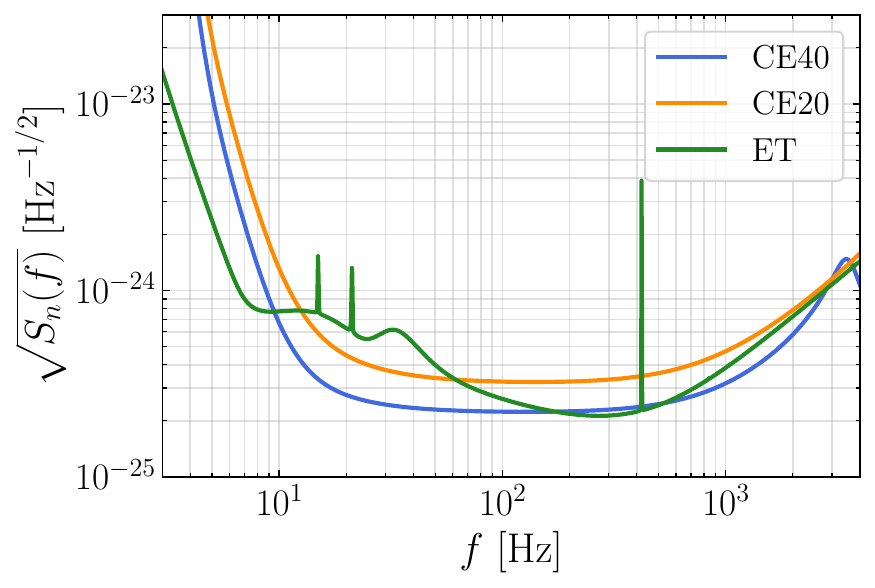}
\caption{Amplitude spectral densities $\sqrt{S_n (f)}$ for the XG detectors we consider. The curve for ET corresponds to the sensitivity of the whole triangular configuration.}
\label{fig:psds}
\end{figure}

\subsection{Data analysis}
\label{sec:fisher}

Let us consider a standard data stream
$d(t) = h({\boldsymbol \theta};t) + n(t)$
composed of Gaussian, stationary noise $n(t)$ and a GW signal $h({\boldsymbol  \theta};t)$ with parameters ${\boldsymbol  \theta}$. To assess whether a signal is detectable, we calculate the optimal SNR
\begin{equation}
    \mathrm{SNR}=\sqrt{\left( h(f)|h(f)\right)}\,,
    \label{eq_SNR}
\end{equation}
where $\left(\cdot|\cdot\right)$ stands for the noise-weighted inner product, defined as
\begin{equation}
\left( a|b\right) = 4\,\mathrm{Re} \int_{f_{\rm min}}^{f_{\rm max}}\frac{\tilde{a}(f)\tilde{b}^*(f)}{S_n(f)}\, {\rm d}f \,.
\end{equation}
Here, $S_n(f)$ is the one-sided power spectral density (PSD) associated with the instrumental noise $n(t)$. We denote Fourier transforms with a tilde and complex conjugates with an asterisk. 

Since we explore the parameter space using a dense grid of $\mathcal{O}(10^6)$ points, performing full Bayesian PE runs for each of the resolved points is computationally unfeasible. Therefore, we estimate the errors within the linear signal approximation~\cite{Finn:1992wt,Vallisneri:2007ev}. In Appendix~\ref{app:pe_comparison} we validate our Fisher-matrix framework against full PE results for a few selected cases.

We model the posterior probabilities for the parameters of each source as a multivariate normal distribution centered at the true values. The covariance matrix is given by the inverse of the Fisher matrix~\cite{Finn:1992wt}
\begin{equation}
\Gamma_{\alpha\beta} = \left(\frac{\del h}{\del\theta^\alpha} \middle| \frac{\del h}{\del\theta^\beta}\right) \,,
\end{equation}
thus the $1\sigma$ error on the parameter $\theta_{\alpha}$ can be estimated as
\begin{equation}
\Delta\theta_\alpha = \sqrt{(\Gamma^{-1})_{\alpha\alpha}}.
\end{equation}
We compute the Fisher matrices on the full set of 11 parameters at our disposal
\begin{eqnarray}
& \Biggl\{ & \ln\left(\frac{\mathcal{M}_z}{M_\odot}\right), 
\eta, 
\chi_{1,z},
\chi_{2,z},
\ln\left(\frac{D_L}{\text{Mpc}}\right), \nonumber \\
&& \cos\iota, 
\cos{\delta},
\alpha, 
\psi,
\phi_{c}, 
t_{c} 
\biggr\} \,,
\label{eq_info_paras}
\end{eqnarray}
where $\mathcal{M}_z = \mathcal{M}_c (1+z)$ is the detector-frame chirp mass, with $\mathcal{M}_c=(m_1m_2)^{3/5}/(m_1+m_2)^{1/5}$, $D_{\rm L}$ is the luminosity distance, and $\eta=(m_1m_2)/(m_1+m_2)^2$ is the symmetric mass ratio. Note that although we inject zero spins for all the events, we include the aligned spin components when computing Fisher matrices. In other words, we do allow for nonzero spins in the recovery of the parameters, capturing the degeneracies between spin magnitudes and other binary parameters (in particular, the mass ratio~\cite{Cutler:1994ys,Hannam:2013uu}). Both SNR and Fisher matrices are computed using the public \texttt{python} package \texttt{GWBENCH}~\cite{Borhanian:2020ypi}. 

Assuming the noise to be uncorrelated among different detectors, both SNR and Fisher matrix can be readily generalized for a network of $N_{\rm{det}}$ detectors as \begin{eqnarray}
\mathrm{SNR}^2 &=& \sum_{j=1}^{N_\mathrm{det}}\mathrm{SNR}^2_j\,, \nonumber \\
\Gamma &=& \sum_{j=1}^{N_{\rm det}} \Gamma_j\,.
\label{eqs_detenet}
\end{eqnarray}

In order to be agnostic on the specific sky location and orientation of each system, all of our results are averaged over the 4 angular parameters $(\cos\iota,\cos\delta,\alpha,\psi)$. In other terms, we define the angle-averaged SNR and error as
\begin{eqnarray}
    \langle \rm{SNR} \rangle &=& \frac{1}{N} \sum_{i=1}^N \rm{SNR}^i\,, \nonumber \\
    \langle \Delta\theta_\alpha \rangle &=& \frac{1}{N} \sum_{i=1}^N \Delta\theta_\alpha^i\,,
\end{eqnarray}
where the sums are conducted over our isotropic grid of points $\{\cos\iota^{(i)},\cos\delta^{(i)},\alpha^{(i)},\psi^{(i)}\}$ for each system. We assume a GW signal to be detected if its network angle-averaged SNR is above the threshold of $\langle \rm{SNR} \rangle = 10$.

\begin{figure*}
\centering
\includegraphics[width=\linewidth]{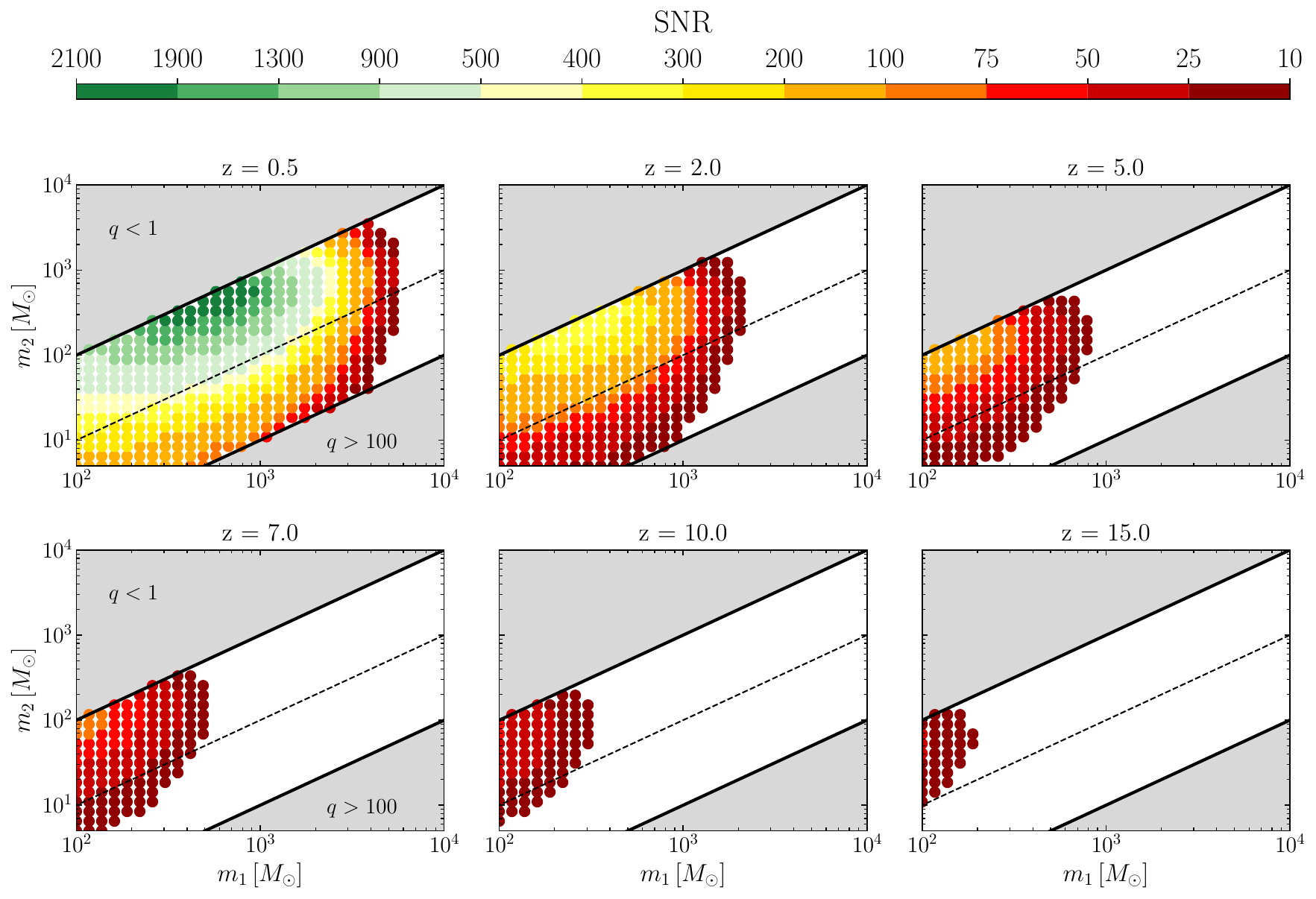}
\caption{Angle-averaged SNR distribution as a function of source-frame masses and redshift for the optimal network (CE40-CE20-ET) with a low-frequency cutoff $f_{\rm min}=3\,\rm{Hz}$. Each panel corresponds to a different redshift and the dots represent detected binaries, colored according to their SNR. The dashed black line corresponds to $q=10$.  The shaded areas correspond to regions of mass ratio $q<1$, which is unphysical, and $q>100$, where our waveform model is unreliable. 
}
\label{fig:avgSNR}
\end{figure*}

\subsection{Detector networks and waveform model}
\label{sec:networks}

In this work, we compare the results with several different networks of XG detectors. Each network is a combination of at least two of the following detectors: a CE with $40~\rm{km}$ arm length (CE40), a CE with $20~\rm{km}$ arm length (CE20), and ET in the triangular configuration~\cite{Branchesi:2023mws}. The PSDs of the detectors are plotted in Fig.~\ref{fig:psds}. The PSDs, locations and orientations of the two CE interferometers correspond to the CE-A and CE-B configurations of Ref.~\cite{Gupta:2023lga}, respectively. We assume ET to be at the current location of Virgo in Italy~\cite{VIRGO:2014yos}.

\begin{figure*}
\centering
\includegraphics[width=\linewidth]{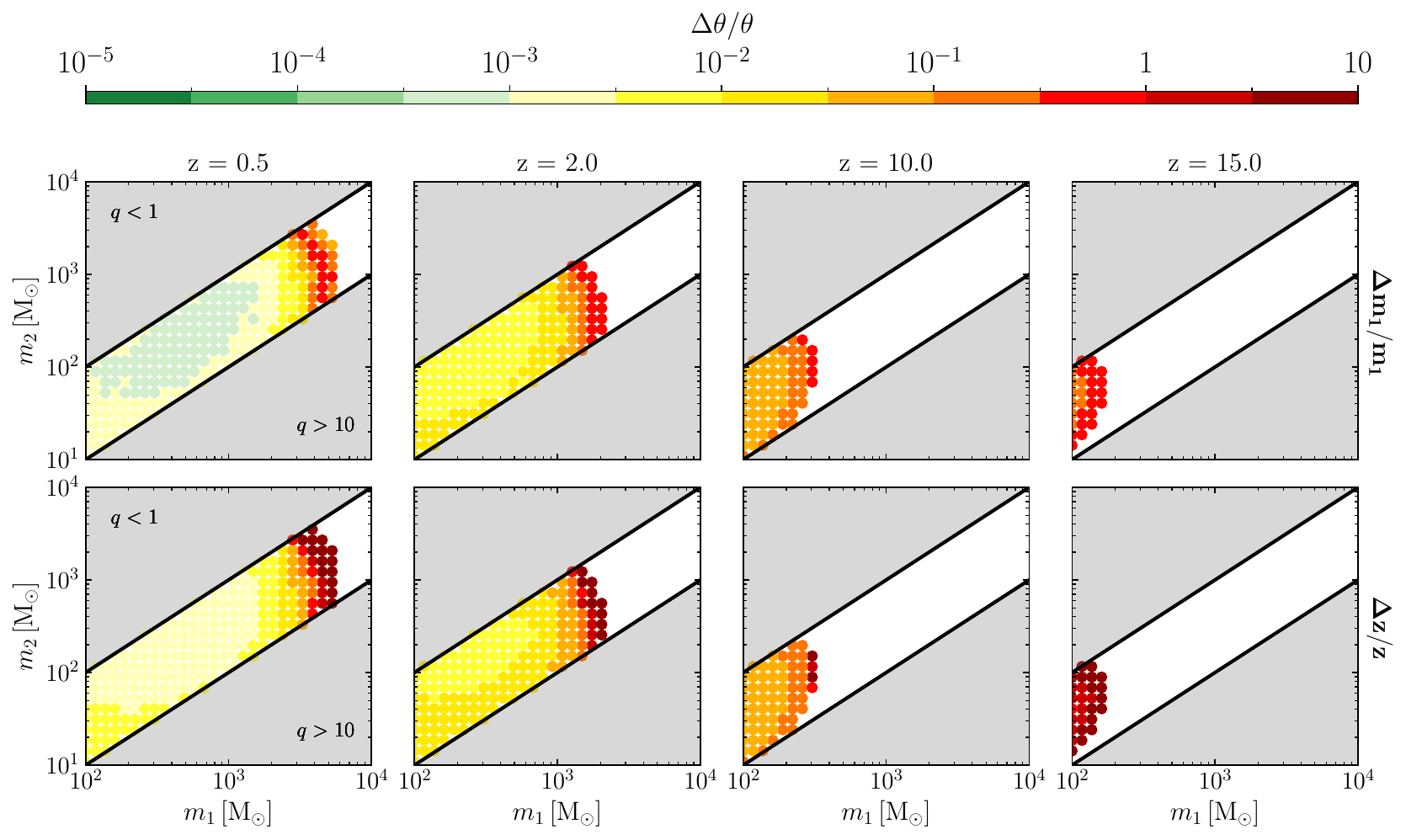}
\caption{Angle-averaged relative errors as a function of source-frame masses and redshift for the optimal network (CE40-CE20-ET) with a low-frequency cutoff $f_{\rm min}=3\,\rm{Hz}$. Panels in the first row show the relative errors on the primary (source-frame) mass, while panels in the second row show the relative error on redshift. The shaded regions correspond to mass ratios $q<1$ and $q>10$, respectively.
}
\label{fig:avgerrors}
\end{figure*}

We consider four detector networks:
\begin{itemize}
    \item An \emph{optimal} network with three XG detectors (CE40-CE20-ET). This corresponds to the best-case scenario~\cite{mpsac_report} and it is identical to the network of 3 XG detectors used in Ref.~\cite{Gupta:2023lga}.
    \item Three comparison networks with two XG detectors, either a combination of one CE and ET (CE40-ET or CE20-ET, respectively) or a combination of two CE interferometers without ET (CE40-CE20). These networks can be compared to the networks with 2 XG detectors of Ref.~\cite{Gupta:2023lga}. (Technically, the networks with 2 XG detectors in Ref.~\cite{Gupta:2023lga} include also some of the current detectors in their $\rm{A}^\#$ configuration. However these detectors are much less sensitive at low frequencies, so we expect their contribution to be minimal for the IMBH binaries considered here.)
\end{itemize}
In Appendix~\ref{app:singledet} we also show results with a single XG detector, either CE40 or ET alone.

The low-frequency reach of ground-based detectors is crucial to detect and characterize IMBH binaries. For a nonrotating BH of mass $M$, the GW frequency at the innermost stable circular orbit (ISCO) is given by~\cite{Maggiore:2007ulw}
\begin{equation}
    f_{\rm ISCO} \approx 4.4\,{\rm{Hz}} \left(\frac{1000\,M_\odot}{M}\right)\,.
\end{equation}
This implies that most IMBH systems considered in this study are expected to merge at GW frequencies $\lesssim 20\,\rm{Hz}$. 
Forecast studies for XG detectors often assume a low-frequency sensitivity cutoff $f_{\rm min}=5\,\rm{Hz}$ for CE~\cite{Gupta:2023lga}, and even $f_{\rm min}=2\,\rm{Hz}$ for ET~\cite{Iacovelli:2022bbs,Wu:2022pyg}. However, these numbers might be optimistic, as they depend on the specific technology adopted (see e.g. the discussion in Ref.~\cite{Branchesi:2023mws}). Moreover, correlated Newtonian noise might be a serious impediment at frequencies $\lesssim 5\,\rm{Hz}$~\cite{Janssens:2022xmo,Janssens:2024jln}. To take into account these uncertainties, when we compare different networks we investigate the impact of the low-frequency sensitivity by considering three different values of $f_{\rm min}$, namely $f_{\rm min}=3$, $7$, and $10\,\rm{Hz}$. 

Throughout our study, we use the \texttt{IMRPhenomXHM} waveform model~\cite{Garcia-Quiros:2020qpx} to generate the GW signals, as it includes higher-order modes that are expected to be important for the detectability and parameter estimation of binaries with IMBH components~\cite{Chandra:2020ccy,Fairhurst:2023beb}. The \texttt{IMRPhenomXHM} model is limited to BH spins aligned with the orbital angular momentum of the binary, and it does not account for spin precession. 
However, even in XG detectors, GWs from IMBH binaries spend only a few cycles in band, making precessional effects hard to detect~\cite{Green:2020ptm,Fairhurst:2023beb}. For this reason, the impact of spin precession on the errors on the parameters of interest in this paper (source-frame masses, redshift, and sky localization) is likely to be subdominant for most IMBH systems. We thus expect that including spin precession will not dramatically change our results, and we leave a detailed exploration of these effects to future studies.

\begin{figure*}
\centering
\includegraphics[width=\linewidth]{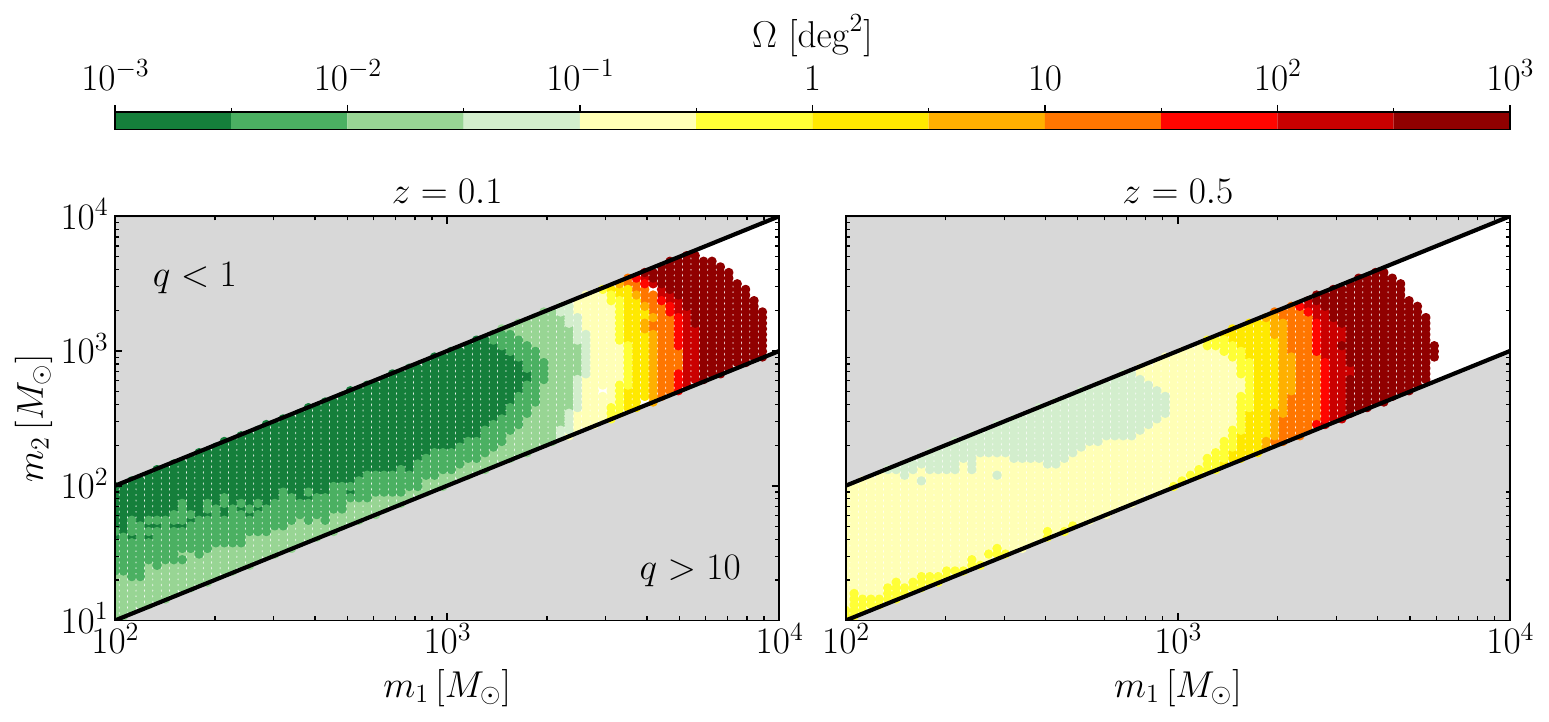}
\caption{Angle-averaged $90\%$ sky localization errors as a function of component masses. The left panel shows the sky localization at $z=0.1$, while the right panel shows the sky localization at $z=0.5$. 
}
\label{fig:avgsky}
\end{figure*}

\section{Results \label{sec:results}}

In this section, we present our results. In Sec.~\ref{sec:optimal} we show SNR and error estimates for the ideal case, i.e., for an optimal CE40-CE20-ET detector network with $f_{\rm min}=3\,\rm{Hz}$. In Sec.~\ref{sec:comparison} we show how the results are affected by less optimistic choices of the detector network and of the low-frequency cutoff.

\subsection{Optimal detector network}
\label{sec:optimal}

In Fig.~\ref{fig:avgSNR} we show the angle-averaged SNR distribution over our parameter space for the resolved sources assuming our optimal configuration: the CE40-CE20-ET network with a low-frequency cutoff $f_{\rm min}=3\,\rm{Hz}$.
The SNR is plotted as a function of source-frame masses $m_{1,2}$ for different redshifts. As expected, sources at lower redshifts have a higher chance of being detected.
Effectively, any binary in our grid with both component masses $\lesssim 1000~\rm{M_\odot}$ can be detected up to redshift $z\sim 2$, with the limiting factors being that sources with higher masses would merge outside of the sensitivity band of the network.
In general, the detectors are more sensitive to comparable-mass systems. In particular, at redshift $z=0.5$ ($z=2$) the SNR distribution shows a peak for comparable-mass binaries in the range $m_{1,2} =[500, 1000]\,M_\odot$ ($m_{1,2} =[100, 800]\,M_\odot$), which can be detected with SNR of $\mathcal{O}(1000)$ ($\mathcal{O}(300)$). The detectability degrades with redshift, but binaries with component masses $m_{1,2} \sim 100\,M_\odot$ can be detected at $z=7$ with $\rm{SNR}\sim 80$, and the bottom left corner of the parameter space is still observable at $z=15$.

For binaries that are detectable, in Fig.~\ref{fig:avgerrors} we plot the angle-averaged relative error distribution on the source-frame primary mass (top row) and redshift (bottom row). We show only the relative errors on primary mass since the errors on secondary mass are qualitatively similar. As discussed in Sec.~\ref{sec:grid}, we further cut the parameter space to exclude regions with $q>10$, as current waveform models are less reliable for higher mass ratios.

We find that, if a binary is detected, the source-frame mass can be confidently constrained up to very high redshift by our optimal network. In particular, for systems with component masses of $\mathcal{O}(1000)\,M_\odot$ or less, the primary mass can be measured with $0.1\%$ accuracy or better at $z=0.5$, and $1\%$ accuracy or better at $z=2$. As the redshift increases, fewer binaries can be observed and the errors get larger. However, it is possible to reach $\mathcal{O}(10)\%$ accuracy for $m_{1,2} \lesssim 300\,M_\odot$, and the relative uncertainty is still below $50\%$ for systems with $m_1 \sim 100\,M_\odot$ that are detectable at redshift $z=15$. These results showcase the promise of XG detectors to provide conclusive evidence of the existence of IMBHs with $\mathcal{O}(1000)\,M_\odot$ and to use them for population studies (see e.g. Ref.~\cite{Franciolini:2024vis}).

\begin{figure*}[hbt]
\centering
\includegraphics[width=\linewidth]{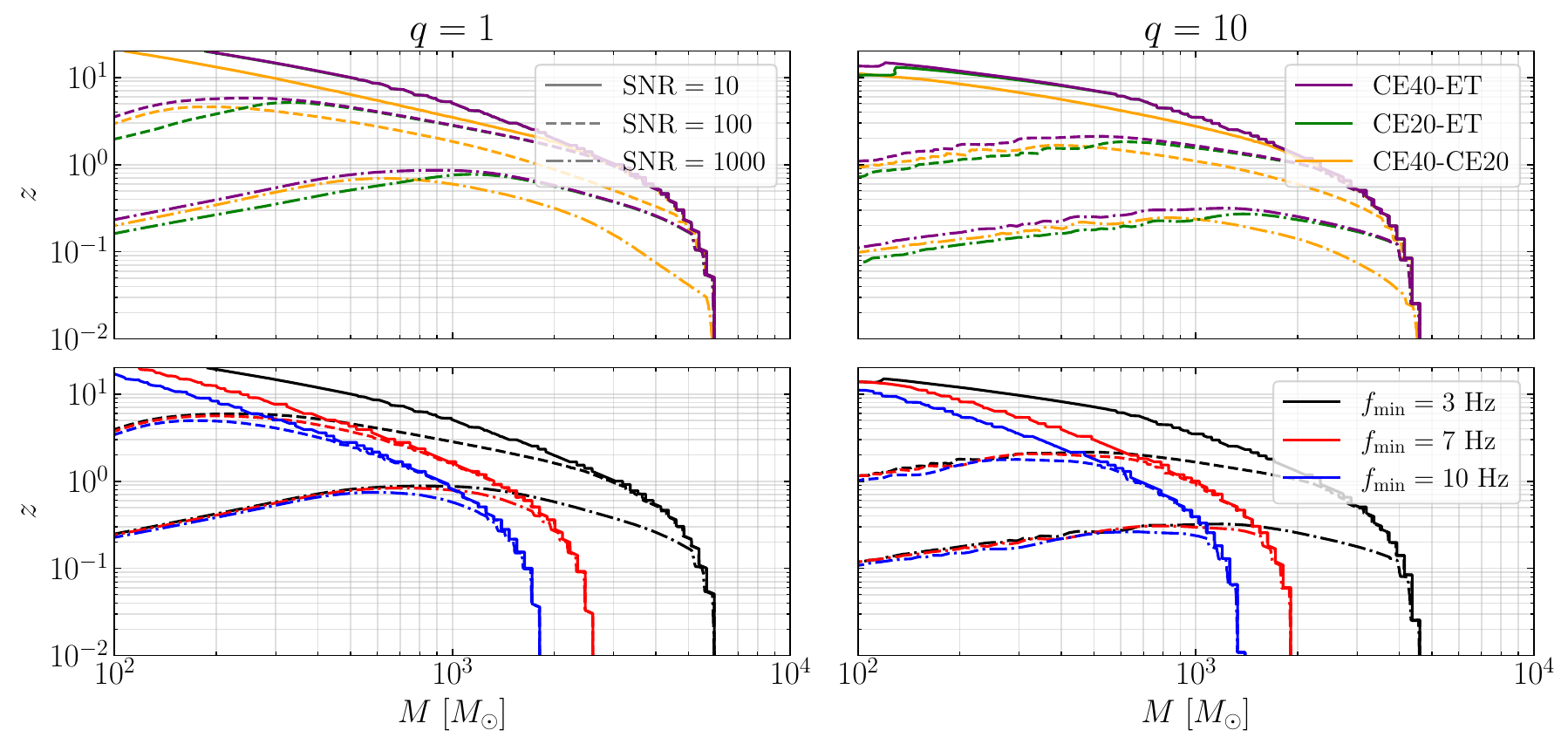}
 \caption{Angle-averaged horizon reach as a function of (source-frame) total mass. The different line styles correspond to different SNR thresholds: 10, 100, and 1000. The left and right panels correspond to different mass ratios ($q=1$ and $q=10$, respectively). The top row shows the horizon reach for different XG networks, shown with different colors, as indicated in the legend. The bottom row shows the horizon for the optimal network with 3 XG detectors, but now different colors correspond to different low-frequency cutoffs, as indicated in the legend. }
\label{fig:snr_comp}
\end{figure*}

The errors on both masses and redshift get larger for asymmetric binaries, showing the expected correlation with the SNR distribution. Furthermore, the errors get larger at the high-mass end of the detectable parameter space, since for those systems only the merger and post-merger phases of the signal are in band. Overall, redshift measurements are less accurate than the measurements of source-frame masses. Redshift is typically measured to an exquisite precision of $\mathcal{O}(0.1)\%$ at $z=0.5$, and to percent level at $z=2$. However, we have uncertainties of order $\sim 10\%$ at $z=10$, while the redshift is essentially unconstrained at $z=15$.

The ability to precisely infer binary parameters can also crucially be exploited to probe cosmology~\cite{Borghi:2023opd,Chen:2024gdn}. IMBH mergers can produce an electromagnetic counterpart if they occur in a gaseous environment~\cite{Bartos:2016dgn,Stone:2016wzz,Graham:2020gwr}, potentially allowing for bright-siren measurements of the Hubble constant~\cite{Chen:2017rfc,Feeney:2018mkj}. Moreover, an accurate 3D localization of binaries without a counterpart could allow us to identify the host galaxy~\cite{Chen:2016tys} or to perform cosmological measurements via correlations with galaxy catalogs~\cite{Mancarella:2022cnu,Borhanian:2020vyr}. In Fig.~\ref{fig:avgsky} we show the distribution of the angle-averaged $90\%$ sky localization error at redshifts $z=0.1$ and $z=0.5$. With our optimal network, we can constrain the sky location to high precision. At redshift $z=0.1$ ($0.5$), effectively all the points with $m_{1,2}\lesssim 3000\,M_\odot$ ($\lesssim 1500\,M_\odot$) have sky areas $\Omega \lesssim 1\,\rm{deg^2}$. The errors get significantly worse at the high-end of the mass spectrum, as signals become too short to allow for an accurate localization of the source. Furthermore, at $z=0.1$ all systems with $m_{1,2}\lesssim 2000\,M_\odot$ are exceptionally well localized, with uncertainties $\lesssim 0.1\,\rm{deg}^2$. In particular, for binaries with $m_{1,2}\lesssim 1000\,M_\odot$ and mass ratios $q < 4$ we can reach sky localizations $\lesssim 0.01\,\rm{deg}^2$. The localization accuracy is worse at $z=0.5$, but there is still a bulk of comparable-mass systems with masses in the range $[200, 700]~\rm{M_\odot}$ that can be localized with uncertainties less than $0.1\,\rm{deg}^2$.

\begin{figure*}[ht]
\centering
\includegraphics[width=\linewidth]{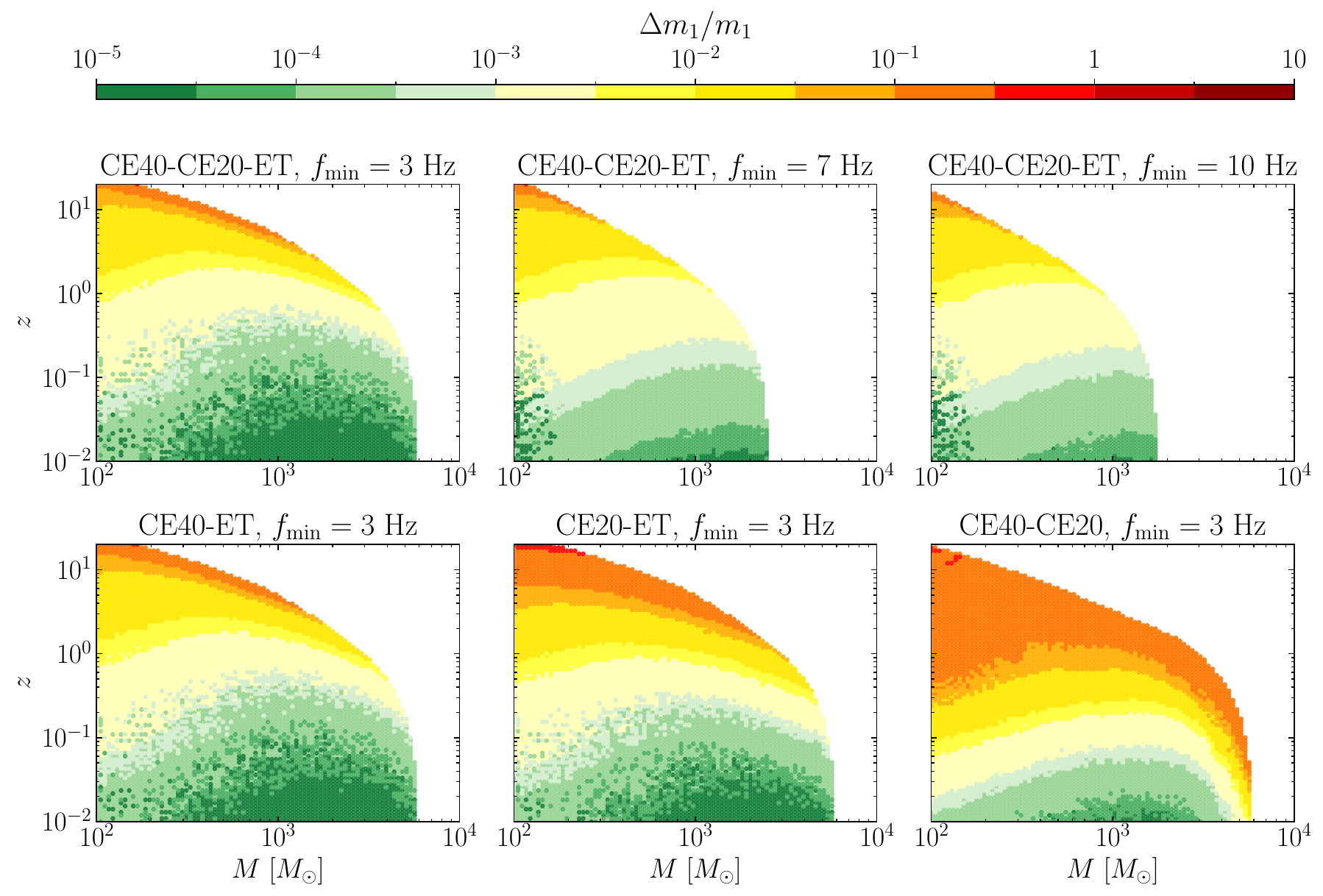}
 \caption{Angle-averaged relative errors on source-frame primary mass as a function of total mass and redshift, for systems with $q=1$. Panels in the first row show the impact of a higher low-frequency cutoff on the optimal network CE40-CE20-ET. Panels in the second row keep the low-frequency cutoff fixed at $f_{\rm min}=3\,\rm{Hz}$ but show results for our three comparison networks.}
\label{fig:deltam1_comp}
\end{figure*}

\subsection{Network comparison}
\label{sec:comparison}

In this section, we assess how much the SNR and parameter estimation errors from the optimal XG network considered above are affected by suboptimal choices of the network and of the low-frequency sensitivity. As discussed in Sec.~\ref{sec:grid}, in this section we switch to a grid in source-frame total mass and redshift, and for computational reasons, we perform the study only for two selected values of the mass ratio ($q=1$ and $q=10$).

In Fig.~\ref{fig:snr_comp} we show the horizon reach for three selected values of the SNR threshold (${\rm SNR}=10$, $100$ and $1000$, shown with different line styles) as a function of the total mass of the binary. The first row compares three networks consisting of 2 XG interferometers: CE40-ET, CE20-ET and CE40-CE20. In the second row we consider again the optimal network CE40-CE20-ET studied in the previous section, but we now change the low-frequency cutoff from $f_{\rm min}=3\,\rm{Hz}$ to $f_{\rm min}=7\,\rm{Hz}$ and $f_{\rm min}=10\,\rm{Hz}$. Under our assumptions, ET is the most sensitive detector at frequencies $\lesssim 10\,\rm{Hz}$, which are crucial for high-mass sources. Both CE interferometers are instead more sensitive in the frequency range $[10, 100]\,\rm{Hz}$ (with CE40 performing better), which is more important at the lower end of the mass spectrum considered here. As a consequence, the CE40-ET network (purple lines in the first row) is remarkably similar to the optimal CE40-CE20-ET network (black lines in the second row) in terms of detectability. Substituting CE40 with CE20 (green lines in the top panels) produces a degradation in the performance at lower masses. In the equal-mass case, the CE20-ET network (green curve) has the lowest reach for $\rm{SNR}=100$ ($1000$) when $M\lesssim 250\,M_\odot$ ($800\,M_\odot$), while it is comparable to CE40-ET for higher masses. The CE20-ET network is comparable to CE40-ET at all values of $M$ for a threshold $\rm{SNR}=10$ (i.e., at high redshift), where the contribution of ET dominates. For the same reason, the CE40-CE20 network has the lowest reach for $\rm{SNR}=100$ ($1000$) when $M\gtrsim 250\,M_\odot$ ($800\,M_\odot$), as well as the worst performance in terms of detectability reach $(\rm{SNR}=10)$ for most values of $M$.

\begin{figure*}[t]
\centering
\includegraphics[width=\linewidth]{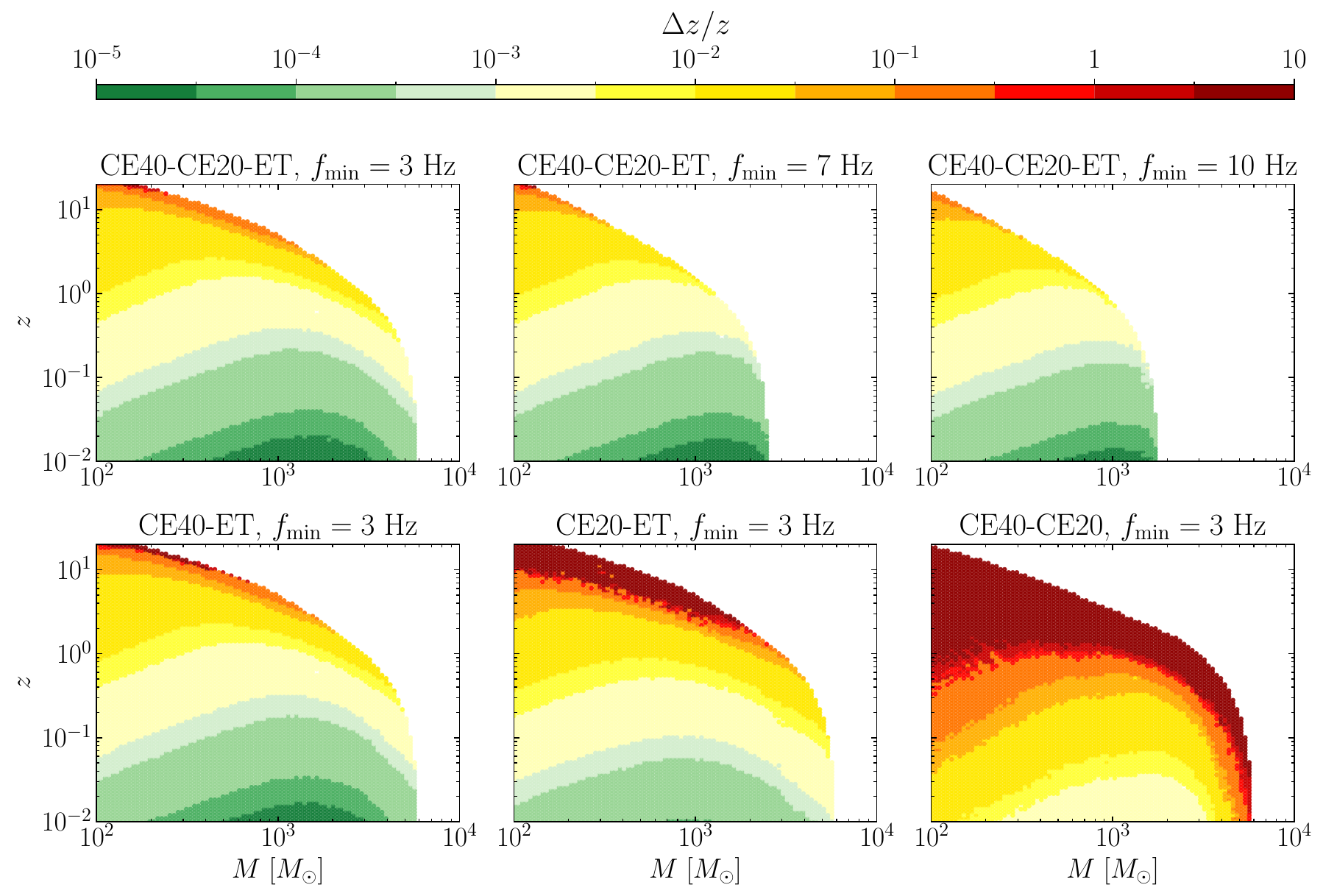}
 \caption{Same as Fig.~\ref{fig:deltam1_comp}, but for angle-averaged relative errors on redshift.}
\label{fig:deltaz_comp}
\end{figure*}

In the second row of Fig.~\ref{fig:snr_comp} we see that the largest impact in terms of detectability is caused by a higher low-frequency cutoff. Even for the optimal network CE40-CE20-ET, the detectability reach corresponding to $\rm{SNR}=10$ for a $M\sim 500\,M_\odot$ ($1000\,M_\odot$), $q=1$ system decreases quite dramatically from $z\sim 10$ ($5$) for $f_{\rm min}=3\,\rm{Hz}$, to $z\sim 4$ ($2$) for $f_{\rm min}=7\,\rm{Hz}$ and $z\sim 1.8$ ($0.8$) for $f_{\rm min}=10\,\rm{Hz}$. Furthermore, networks with $f_{\rm min}=7\,\rm{Hz}$ and $f_{\rm min}=10\,\rm{Hz}$ cannot detect any sources above $\sim 2000\,M_\odot$ and $\sim 3000\,M_\odot$, respectively, even at the lowest redshifts. However, the horizons at $\rm{SNR}=100$ ($1000$) are fairly similar for all three values of $f_{\rm min}$ below masses of $200\,M_\odot$ ($600\,M_\odot$). The implication is that it is still possible to have high-precision IMBH parameter inference even for $f_{\rm min}=10\,\rm{Hz}$, at least at the lower end of the mass spectrum.

Having assessed the performance of the different networks in terms of detectability, we now compare the measurement errors on various parameters.

In Fig.~\ref{fig:deltam1_comp} we show the angle-averaged relative errors on source-frame primary mass, $\Delta m_1/m_1$, for different networks and different choices of the low-frequency cutoff. As before, we display only the errors on the primary mass, because the errors on the secondary mass are qualitatively similar. We focus on $q=1$ binaries, for which we can typically get better constraints (see Fig.~\ref{fig:avgerrors}). 

In the top row of Fig.~\ref{fig:deltam1_comp} we show how much the errors are affected by a degradation of the low-frequency sensitivity in the optimal network (CE40-CE20-ET). The parameter space for which masses can be measured shrinks significantly as $f_{\rm min}$ increases: consistently with Fig.~\ref{fig:snr_comp}, the detectability threshold moves to lower masses and lower redshift as systems with high detector-frame mass start merging outside of the network sensitivity band. For instance, a binary with total mass $M\sim1000\,M_\odot$ at redshift $z\sim 2$ would be observable with $<10\%$ uncertainty in $m_1$ when $f_{\rm min}=3\,\rm{Hz}$, while it would not be detected for $f_{\rm min}=7$ or $f_{\rm min}=10\,\rm{Hz}$. The maximum mass that can be constrained shifts from $\sim 6000\,M_\odot$ for $f_{\rm min}=3\,\rm{Hz}$, to $\sim 2500\,M_\odot$ for $f_{\rm min}=7\,\rm{Hz}$ and $\sim 1800\,M_\odot$ for $f_{\rm min}=10\,\rm{Hz}$. For systems with $M\sim 100\,M_\odot$, the horizon redshift at which $\Delta m_1/m_1<0.1$ is reduced from $z\sim18$ to $z\sim15$ and $z\sim12$ for $f_{\rm min}=3$, $7$ and $10\,\rm{Hz}$, respectively. The region that allows for high-precision measurements of the mass also gets remarkably smaller. For systems with $M\sim 1000\,M_\odot$, the primary mass can be measured with $0.1\%$ ($0.01\%$) accuracy or better up to redshifts $z \sim 0.6$ ($0.1$), $\sim 0.3$ ($0.02$) and $\sim 0.2$ ($0.01$) for $f_{\rm min}=3$, $7$, and $10\,\rm{Hz}$, respectively. However, in the central regions of the parameter space the performance does not get much worse. For sources that are detectable for all values of $f_{\rm min}$, the horizon for $1\%$ accuracy on mass is fairly close. Even in the worst-case scenario ($f_{\rm min}=10\,\rm{Hz}$), $m_1$ can still be constrained to subpercent level for binaries with $M\sim 500\,M_\odot$  up to redshift $z\sim 2$.

In the bottom row of Fig.~\ref{fig:deltam1_comp} we compare different networks with 2 XG interferometers, assuming the ideal low-frequency cutoff $f_{\rm min}=3\,\rm{Hz}$. The key takeaway from these plots is that the contribution of ET is crucial for IMBH binaries due to its higher sensitivity at low frequencies. The CE40-ET performs only slightly worse than the optimal CE40-CE20-ET network. Mass measurements progressively degrade as we move to the CE20-ET and CE40-CE20 networks, with the latter performing the worst. For a binary with total mass $\sim 100\,M_\odot$ ($1000\,M_\odot$), the horizon redshift at which $\Delta m_1/m_1<0.1$ shrinks from $z\sim 16$ ($3$) for the CE40-ET network, to $z\sim 6$ ($2$) for CE20-ET and $z\sim 0.4$ ($1$) for CE40-CE20. We observe similar behavior for the high-precision measurement regions, where the mass $m_1$ for a binary with $M=1000\,M_\odot$ can be constrained with $0.1\%$ ($0.01\%$) accuracy or better up to $z\sim 0.5$ ($0.1$) for the CE40-ET network, $z\sim 0.2$ ($0.06$) for CE20-ET, and $0.06$ ($0.01$) for CE40-CE20.

It is well known that individual GW detectors measure only redshifted masses, while tracing the merger history of IMBHs requires also a measurement of the source redshift.  
In Fig.~\ref{fig:deltaz_comp} we show the relative error $\Delta z/z$ for different choices of the low-frequency sensitivity limit (top row) and different XG detector networks (bottom row). As in Fig.~\ref{fig:avgerrors}, we find that redshift errors are typically larger than mass measurement errors in all cases. As we observed in Fig.~\ref{fig:deltam1_comp}, increasing $f_{\rm min}$ significantly shrinks the available parameter space of sources that merge in band, and whose parameters can be measured. The redshift of an IMBH system with $M\sim 1000\,M_\odot$ can be constrained with $\lesssim 10\%$ precision up to redshift $\sim 3$ for $f_{\rm min}=3\,\rm{Hz}$, while the binary would not be detectable at all at such redshifts when $f_{\rm min}=7$ or $10\,\rm{Hz}$. For observed binaries that are far away from the boundaries of the parameter space, however, the constraints in redshift get only slightly worse with increasing $f_{\rm min}$ (the horizons shown in the three different panels of the top row are fairly close to each other). The most significant degradation happens at high redshifts: for binaries with total mass of order $\mathcal{O}(100)\,M_\odot$, the horizon for which $\Delta z/z\lesssim 0.1$ goes from $z\sim 17$ for $f_{\rm min}=3\,\rm{Hz}$ to $z\sim 15$ for $f_{\rm min}=7\,\rm{Hz}$, and to $z\sim 12$ for $f_{\rm min}=10\,\rm{Hz}$.

\begin{figure*}[t]
\centering
\includegraphics[width=\linewidth]{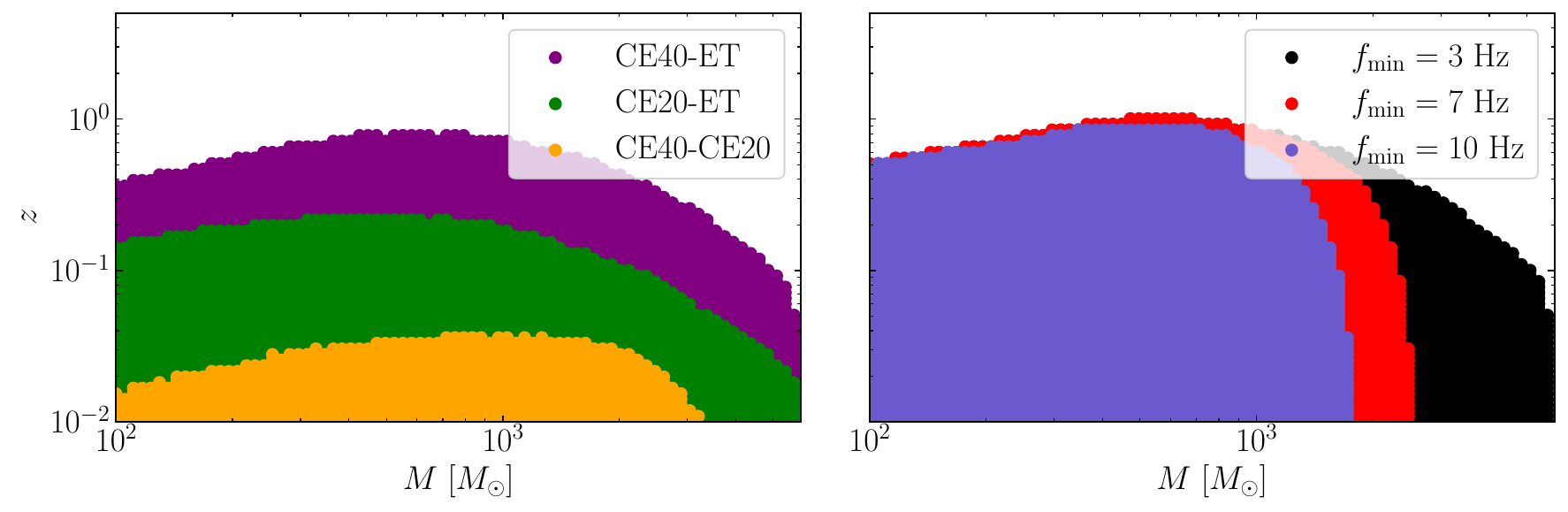}
 \caption{Angle-averaged horizons for which the sky localization is constrained within $1\,\rm{deg}^2$ or better, as functions of the total source-frame mass and redshift. Here we focus on equal-mass binaries ($q=1$). In the left panel, we compare the three networks of 2 XG detectors, while in the right panel, we show the effect of increasing the low-frequency cutoff on our optimal network.}
\label{fig:skyarea_comp}
\end{figure*}

By comparing different detector networks (bottom row of Fig.~\ref{fig:deltaz_comp}), we find once again that the CE40-ET network performs only slightly worse than the optimal CE40-CE20-ET network. Redshift errors become larger and larger as we consider the CE20-ET and the CE40-CE20 networks, respectively. In particular, the CE40-ET network is capable of high-precision redshift measurements with accuracies $\lesssim 0.1\%$ up to $z\sim 0.2$ for a $\sim 1000\,M_\odot$ system. This region is reduced to $z\sim 0.09$ for the CE20-ET network, while it disappears completely for CE40-CE20. Even more importantly, the horizon at which the redshift can be confidently measured is significantly affected by the choice of network. The CE40-ET network can infer the redshift of binaries of $M\sim 100\,M_\odot$ ($1000\,M_\odot$) with $\lesssim 10\%$ uncertainty up to $z\sim 15$ ($4$); the CE20-ET network can do so up to $z\sim 5$ ($2.5$), while the CE40-CE20 can reach this precision only up to $z\sim 0.1$ ($0.4$).

Finally, in Fig.~\ref{fig:skyarea_comp} we compare the performance of different XG networks and the effect of different choices of $f_{\rm min}$ in terms of sky localization. Once again, increasing the low-frequency cutoff reduces the reach in mass and redshift, but the performance on sources that are resolved is similar across all values of $f_{\rm min}$. The sky location of a binary of $\sim 3000\,M_\odot$ can be measured within $1\,\rm{deg}^2$ or better up to $z\sim0.4$ with the optimal network and $f_{\rm min}=3\,\rm{Hz}$, while the binary is not detectable at all with the same network and $f_{\rm min}=7$ or $10\,\rm{Hz}$. A system of $\sim 2000\,M_\odot$ can be localized within less than $1\,\rm{deg}^2$ up to $z\sim0.5$ for $f_{\rm min}=3\,\rm{Hz}$ and $z\sim0.4$ for $f_{\rm min}=7\,\rm{Hz}$, but it would not be observed at all if $f_{\rm min}=10\,\rm{Hz}$. However, binaries of $\sim 800\,M_\odot$ ($100\,M_\odot$) can be localized within less than $1\,\rm{deg}^2$ up to at least $z\sim 0.8$ ($0.5$) by the optimal network for all values of $f_{\rm min}$. 

The left panel compares different networks with 2 XG detectors at fixed $f_{\rm min}=3\,\rm{Hz}$. Once again, the performance in the CE40-ET case is only slightly worse than the optimal CE40-CE20-ET network shown in the right panel. For binaries with total mass $\sim 1000\,M_\odot$ ($100\,M_\odot$), the horizon for $\lesssim 1\,\rm{deg}^2$ sky localization goes from $z\sim 0.9$ ($0.5$) with CE40-CE20-ET to $z\sim 0.7$ ($0.4$) with CE40-ET. The constraints are significantly worse for the CE20-ET and (even more so) for the CE40-CE20 network. In the former case, binaries with $M\sim 1000\,M_\odot$ ($100\,M_\odot$) could be localized within $\lesssim 1\,\rm{deg}^2$ up to $z\sim 0.15$; in the latter, the horizon reaches only up to $z\sim 0.03$ and $z\sim 0.015$ for $\sim 1000\,M_\odot$ and $100\,M_\odot$, respectively. This would limit follow-up campaigns only to rare, close-by sources.

\section{Conclusions}
\label{sec:conclusions}

In this work, we have assessed the detectability of binaries with at least one IMBH, and the measurability of their parameters, with XG observatories.

Since IMBH formation mechanisms are highly uncertain, we have performed an agnostic parameter scan by considering a log-uniform grid in masses and a uniform grid in redshift. We have used a Fisher information matrix formalism to estimate errors on source-frame masses, redshift, and sky localization by averaging over sky position, orbital inclination, and polarization angle. We have first computed PE errors as a function of the binary component masses $m_{1,2}$ at different redshifts for an optimal network of 3 XG detectors, assuming a low-frequency sensitivity cutoff of $f_{\rm min}=3\,\rm{Hz}$. We then compared the detectability reach and measurement errors that would result from suboptimal (two-detector) networks and more pessimistic choices of the low-frequency cutoff.

\begin{figure*}[hbt]
\centering
\includegraphics[width=\linewidth]{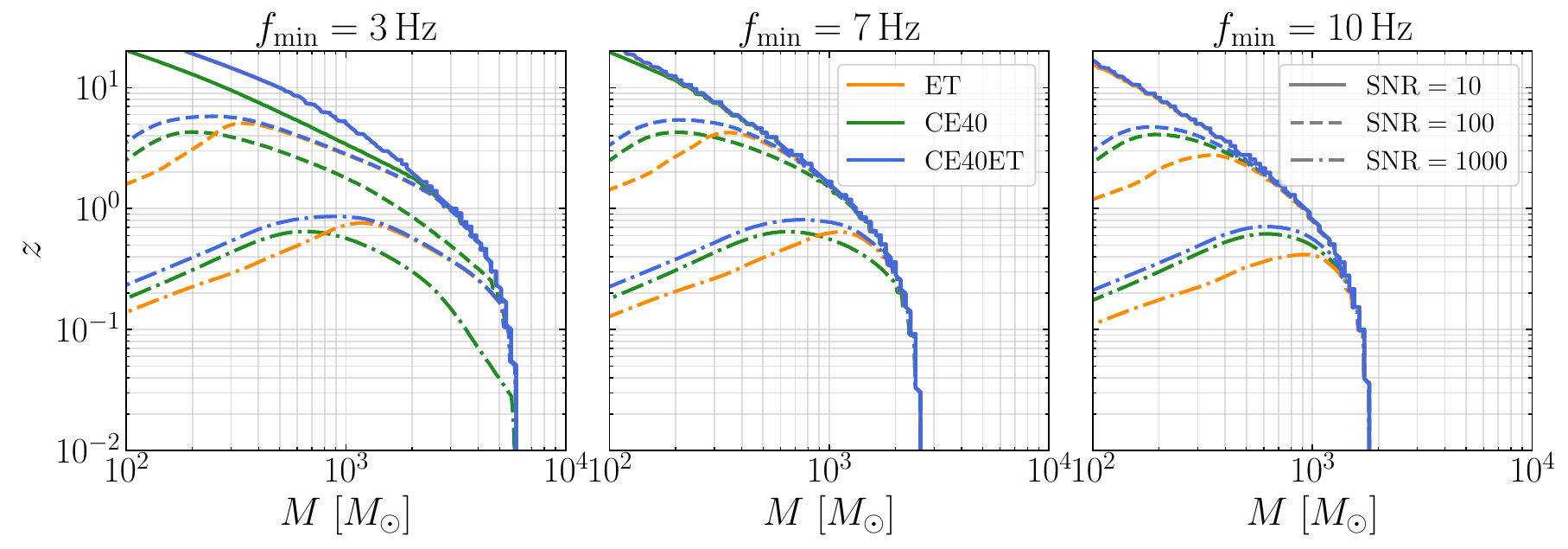}
 \caption{Angle-averaged horizon reach as a function of (source-frame) total mass for binaries with mass ratio $q=1$. Different line styles correspond to different SNR thresholds: 10, 100, and 1000. The colors correspond to a scenario with only CE40 (green), only ET (orange) and both CE40 and ET (blue). The three panels show the impact of different low-frequency sensitivity limits ($f_{\rm min}=3\,\rm{Hz}$, $f_{\rm min}=7\,\rm{Hz}$, and $f_{\rm min}=10\,\rm{Hz}$, from left to right).}
\label{fig:snr_comp_singledet}
\end{figure*}

The optimal network can constrain component masses for systems of $m_{1,2}\sim 1000\,M_\odot$ with $\lesssim 0.1\%$ errors at $z=0.5$, and $\lesssim 1\%$ errors at $z=2$. Such heavy systems merge outside the sensitivity band of XG detectors at high redshifts, but it is still possible to constrain the masses of binaries with $m_{1,2}\lesssim 300\,M_\odot$ with $\lesssim 10\%$ uncertainty at $z=10$. For binaries with component masses $m_{1,2}\sim 1000\,M_\odot$, the redshift can be measured with percent-level accuracy or better at $z=2$, and it can still be measured with $\mathcal{O}(10)\%$ accuracy for $m_{1,2}\lesssim 300\,M_\odot$ binaries at $z=10$. As long as IMBH merger rates are large enough, this suggests that XG detectors could provide a census of the lower end of the IMBH mass spectrum throughout cosmic history, thus shedding light on their population and astrophysical formation scenarios.

Low-redshift binaries with $z\lesssim 0.1$ ($\lesssim 0.5$) can be localized within $0.1\,\rm{deg}^2$ ($1\,\rm{deg}^2$) for $m_{1,2}\lesssim 2000\,M_\odot$ ($\lesssim 1000\,M_\odot$), and within $0.01\,\rm{deg}^2$ ($0.1\,\rm{deg}^2$) for comparable mass systems. The sky localization is good enough that it may be possible to cross-correlate GW searches with galaxy catalogs and search for electromagnetic counterparts to IMBH mergers.

We find that the low-frequency sensitivity of the detectors is crucial for both the detection and PE of IMBH binaries. Degrading the low-frequency cutoff from $3$ to $7$ or $10\,\rm{Hz}$ significantly reduces the reach of the detectors in both mass and redshift. For example, an equal-mass binary with total mass $M\sim 1000\,M_\odot$ at $z=2$ can be detected with percent-level uncertainties in both component masses and redshift for $f_{\rm min}=3\,\rm{Hz}$, while it would not be detectable at all for $f_{\rm min}=7$ or $10\,\rm{Hz}$. However, lower-mass IMBH binaries with $M\sim 500\,M_\odot$ can be observed with subpercent level errors in both component masses and redshift up to $z\sim 2$, even in the worst-case scenario in which $f_{\rm min}=10\,\rm{Hz}$.

The sensitivity of individual detectors also plays an important role.  Among the interferometers we considered, ET is the most sensitive at frequencies $\lesssim 10\,\rm{Hz}$, while CE40 is the most sensitive in the range $[10,100]\,\rm{Hz}$. This implies that ET plays a dominant role for IMBHs with high detector-frame masses, while CE40 is the most valuable detector at the lower-mass end. For this reason, a CE40-ET network is only marginally worse than the optimal CE40-CE20-ET network at constraining masses and redshift. Parameter estimation errors are slightly worse for the CE20-ET network, while a hypothetical network consisting only of two CEs (CE40-CE20) would have the worst performance among networks consisting of 2 XG detectors, constraining IMBH binary masses and redshift with $\lesssim 10\%$ errors only for $z\lesssim 1$. For an IMBH binary with $M\sim 1000\,M_\odot$, the optimal CE40-CE20-ET network can reach sky localization precision below $1\,\rm{deg}^2$ up to $z\sim 0.9$; CE40-ET, up to $z\sim 0.7$; CE20-ET, up to $z\sim0.15$; and CE40-CE20, only up to $z\sim0.015$.

In conclusion, we wish to point out some important caveats. First of all, we modeled the instrumental noise by only taking into account the detector PSD, but the mass and redshift reach for high-mass systems are currently affected by noise glitches that can mimic short signals~\cite{Chandra:2020ccy}. This is likely to be the case for XG detectors as well, especially for signals that merge close to the low-frequency sensitivity limit. For this reason, our horizon estimates for detectable sources are somewhat optimistic. Moreover, the presence of confusion noise due to overlapping signals at low frequencies can further limit the reach of the detectors~\cite{Wu:2022pyg} and slightly worsen the parameter estimation errors of resolved signals~\cite{Reali:2022aps,Reali:2023eug,Johnson:2024foj}. We leave a detailed exploration of these effects to future work. 

\begin{figure*}[hbt]
\centering
\includegraphics[width=\linewidth]{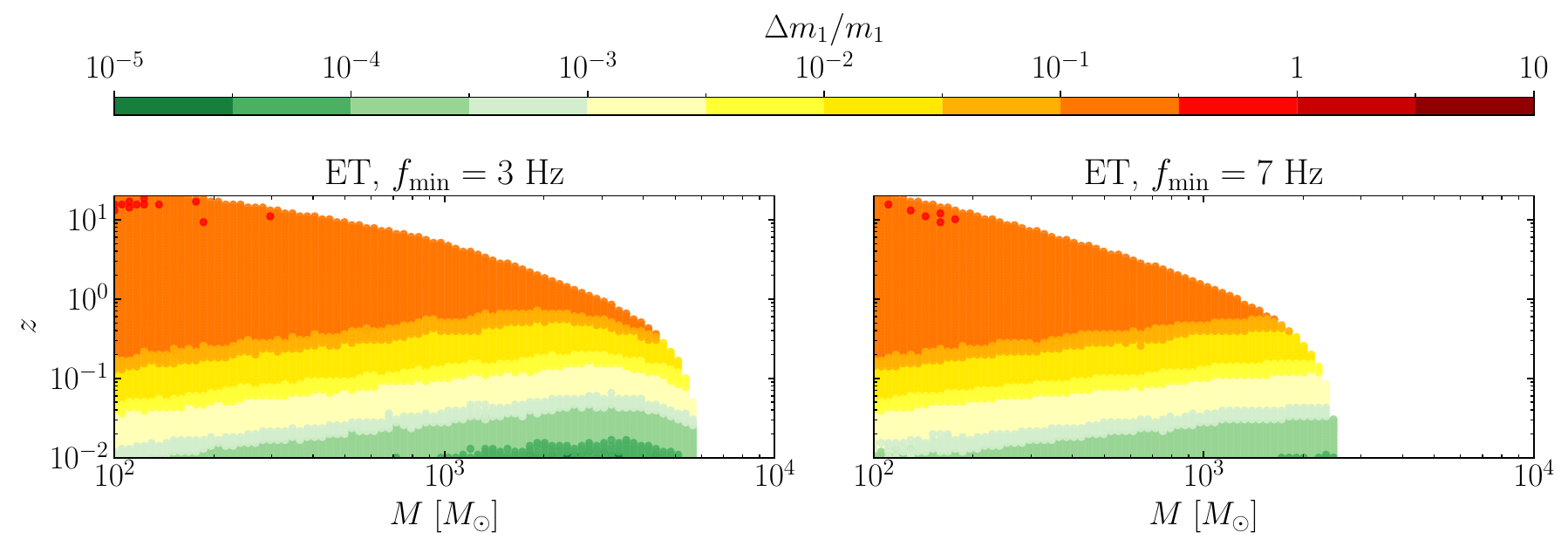}
\includegraphics[width=\linewidth]{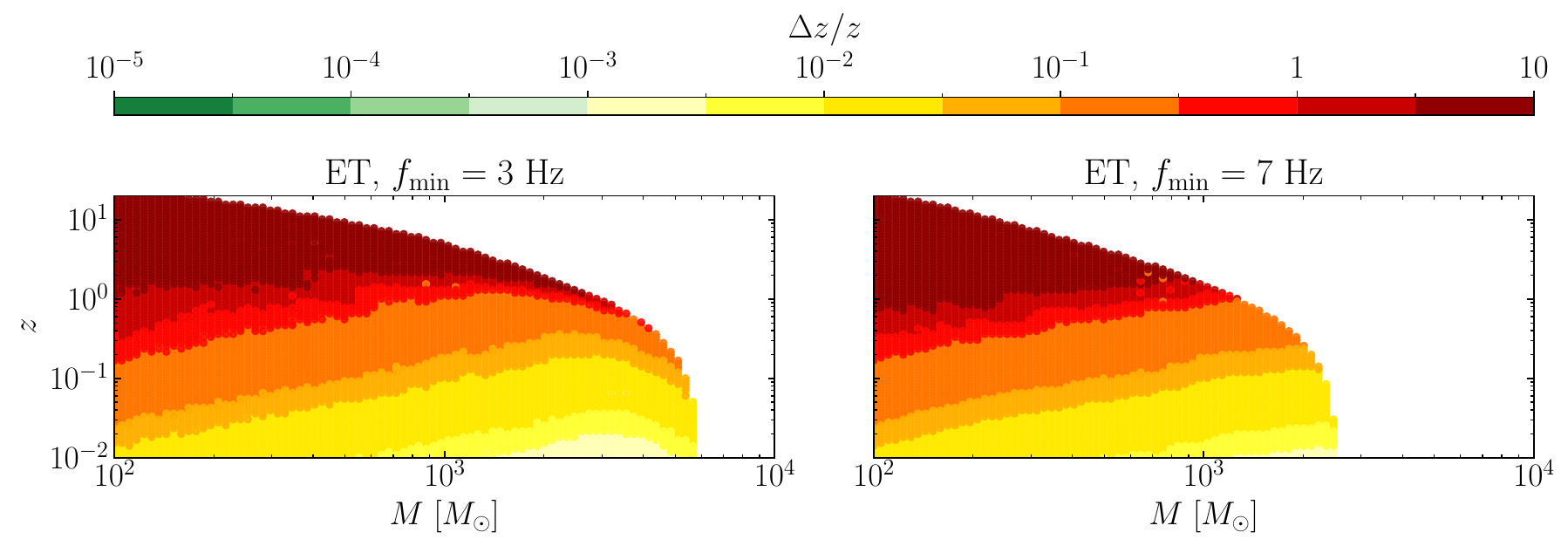}
 \caption{Angle-averaged relative errors on source-frame primary mass (upper panels) and redshift (lower panels) as functions of total mass and redshift for binaries with mass ratio $q=1$. We consider ET only, and set the low-frequency sensitivity limit of the detector to be either $3\,\rm{Hz}$ (left panels) or $7\,\rm{Hz}$ (right panels).}
\label{fig:errors_ET}
\end{figure*}

\acknowledgements

The authors thank the organizers and participants of the 2022 IMBH meeting in Puerto Rico for discussions. 
L.~Reali, R.~Cotesta, A.~Antonelli, K.~Kritos, V.~Strokov, and E.~Berti are supported by NSF Grants No. AST-2006538, PHY-2207502, PHY-090003, and PHY-20043, by NASA Grants No. 20-LPS20-0011 and 21-ATP21-0010, by the John Templeton Foundation Grant 62840, by the Simons Foundation, and by the Italian Ministry of Foreign Affairs and International Cooperation grant No.~PGR01167. 
K.~Kritos is supported by the Onassis Foundation - Scholarship ID: F ZT 041-1/2023-2024.
This work was carried out at the Advanced Research Computing at Hopkins (ARCH) core facility~\cite{rockfish}, which is supported by the NSF Grant No.~OAC-1920103.
The authors acknowledge the Texas Advanced Computing Center (TACC) at The University of Texas at Austin for providing {HPC, visualization, database, or grid} resources that have contributed to the research results reported within this paper \cite{10.1145/3311790.3396656}.

\appendix 

\section{Individual detectors}
\label{app:singledet}

In this Appendix, we briefly discuss how a single XG detector would perform compared to the networks considered in the rest of this study.

In Fig.~\ref{fig:snr_comp_singledet} we compare the angle-averaged horizons at different SNR thresholds between ET alone, CE40 alone, and a CE40-ET network. The main trends are similar to those discussed in Sec.~\ref{sec:comparison}. ET is the more sensitive detector at frequencies $f\lesssim 10\,\rm{Hz}$, hence it performs better than CE40 (and remarkably close to the CE40-ET network) for binaries with high masses and/or at high redshifts. The difference in performance between ET and CE40 in these regions of the parameter space progressively decreases as we increase the low-frequency sensitivity limit from $3\,\rm{Hz}$ to $10\,\rm{Hz}$. Vice versa, CE40 is the more sensitive detector in the frequency range $[10, 100]\,\rm{Hz}$, so it performs better at the lower end of the IMBH mass spectrum.

\begin{figure*}[ht]
\centering
\includegraphics[width=0.49\linewidth]{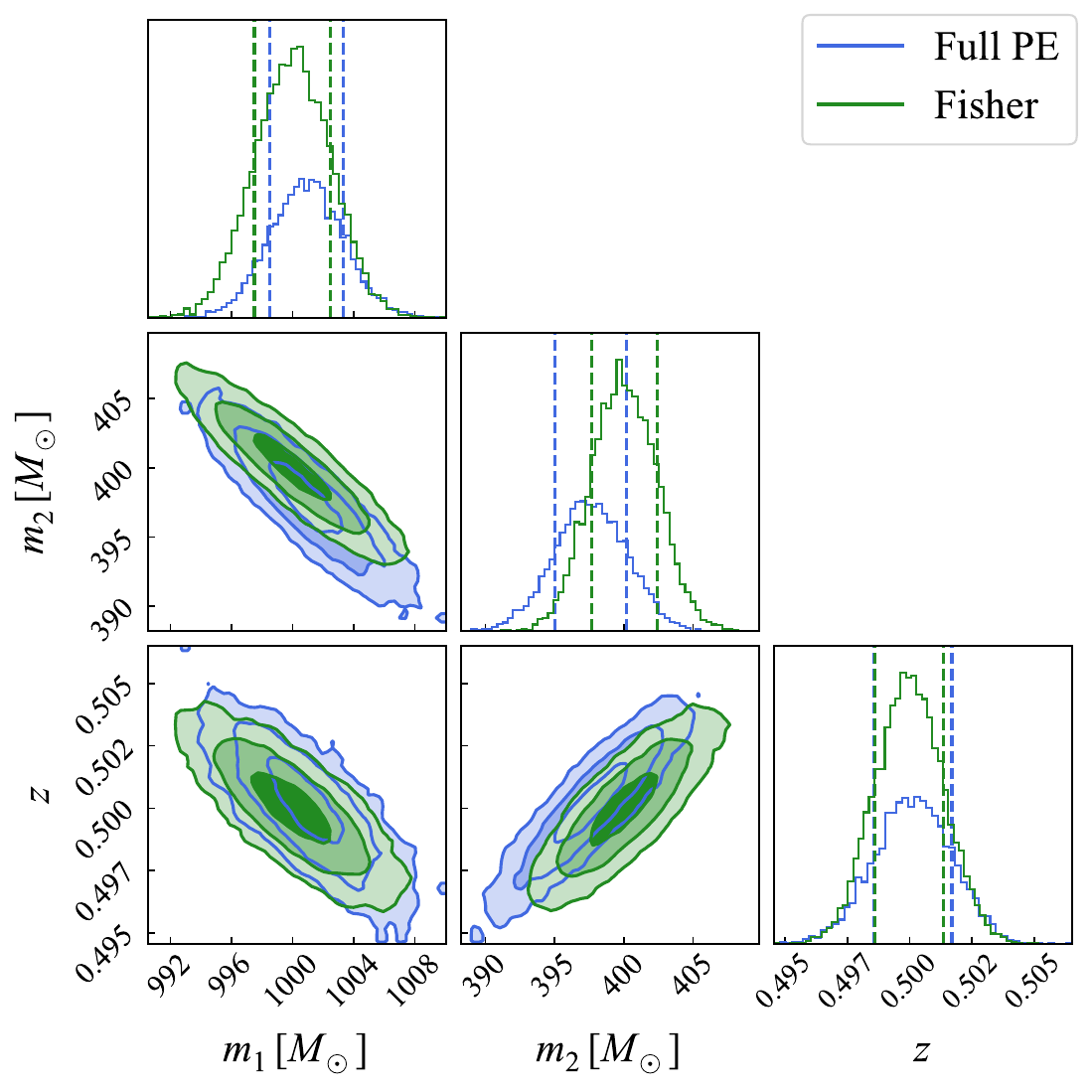}
\includegraphics[width=0.49\linewidth]
{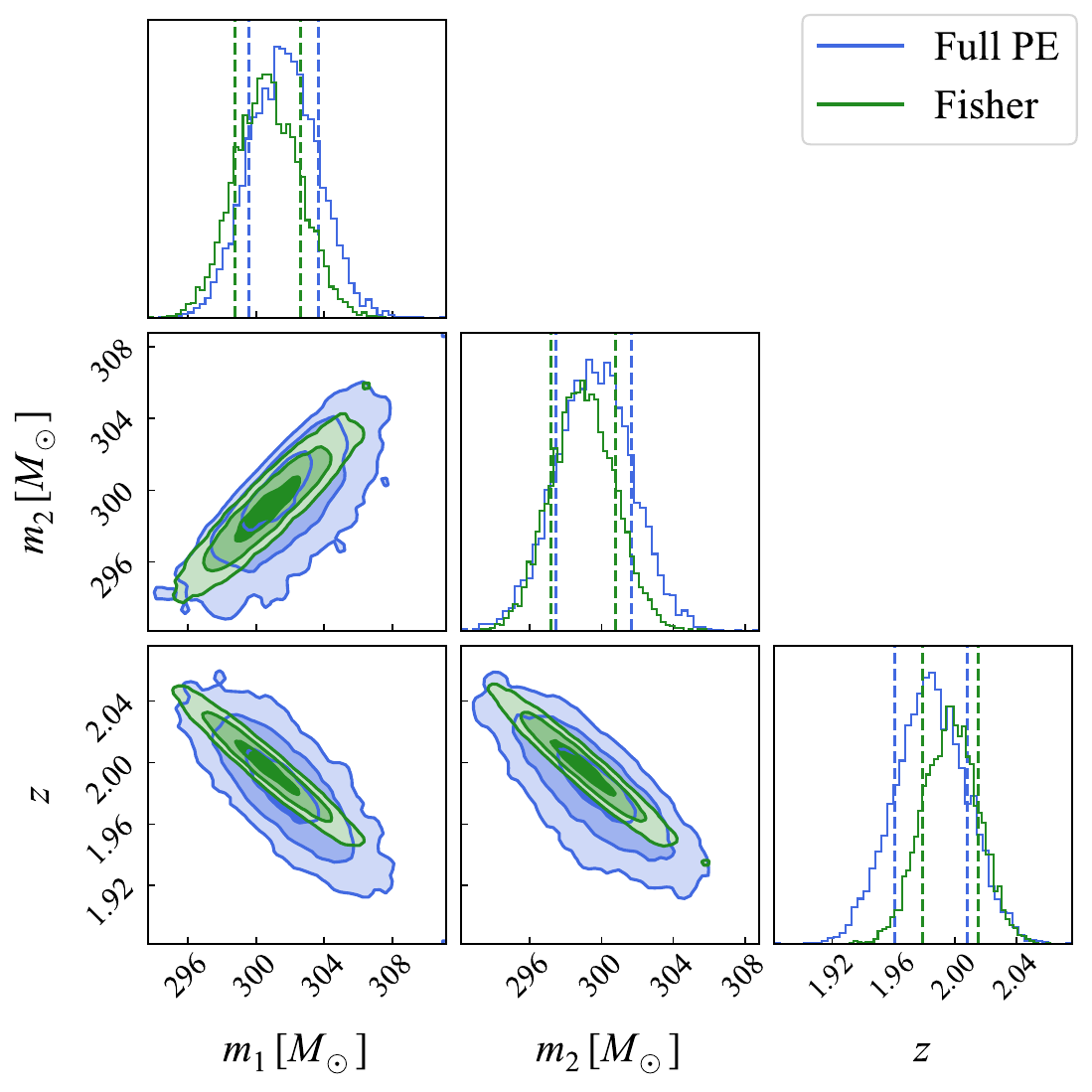}
 \caption{Comparison between posterior distribution obtained with full PE (blue curves) and in the Fisher approximation (green curves) for two representative IMBH binaries. The left panels show results for a system with the following parameters: $m_1=1000\,M_\odot$, $m_2=400\,M_\odot$, $z=0.5$, $\chi_{1,z}=0$, $\chi_{2,z}=0$, $\iota=2.66$, $\alpha=0.01$, $\delta=-0.27$, $\psi=1.86$. The right panels show results for a system with the following parameters:  $m_1=300\,M_\odot$, $m_2=300\,M_\odot$, $z=2$, $\chi_{1,z}=0$, $\chi_{2,z}=0.1$, $\iota=1.02$, $\alpha=3.23$, $\delta=-0.16$, $\psi=1.38$. The 2D contours correspond to $1\sigma$, $2\sigma$ and $3\sigma$ levels, while the dashed vertical lines delimit $1\sigma$ intervals in the marginalized 1D distributions.}
\label{fig:pe_comp}
\end{figure*}

Figure~\ref{fig:errors_ET} is shown as an example of how the measurability of IMBH parameters gets degraded if we consider a single detector (in this case, ET). It can be compared with Figs.~\ref{fig:deltam1_comp} and \ref{fig:deltaz_comp}. The errors on the primary mass and redshift with ET alone are significantly worse than with any network containing both ET and a CE detector (namely, CE40-CE20-ET and CE40-ET) across the entire parameter space that we consider. However, this is not the case if we compare ET with a CE40-CE20 network, as ET provides worse constraints in the majority of the parameter space, but better ones at the higher end of the mass spectrum. For instance, with a low-frequency limit of $3\,\rm{Hz}$, the primary mass of a $1000\,\rm{M}_\odot$ ($5000\,\rm{M}_\odot$) binary can be constrained with percent-level accuracy up to redshift $z\sim 1$ ($z\sim 0.008$) with CE40CE20, and $z\sim 0.5$ ($z\sim 0.4$) with ET. These observations are once again consistent with the fact that ET is the more sensitive detector at low frequencies, while CE detectors perform better in the frequency range $[10, 100]\,\rm{Hz}$.

\section{Comparison with full PE}
\label{app:pe_comparison}

Here we validate our Fisher results by reporting comparisons with full Bayesian parameter estimation (PE) using \texttt{BILBY}~\cite{Ashton:2018jfp}. We perform the comparison for our CE40-CE20-ET network with a low-frequency sensitivity limit of $f_{\rm min}=3\,\rm{Hz}$.

In Fig.~\ref{fig:pe_comp} we show the posterior distributions for two representative IMBH binaries, namely a binary with 
$m_1=1000\,\rm{M_\odot}$, $m_2=400\,\rm{M_\odot}$ at $z=0.5$ (left panels) and a binary with $m_1=300\,\rm{M_\odot}$, $m_2=300\,\rm{M_\odot}$ at $z=2$ (right panels). All BHs have zero spin, while the angular parameters are randomly sampled. The two signals have network SNRs of $\sim 1460$ and $\sim 260$, respectively. We find our Fisher approximation of the posterior to be in good agreement with the full PE results. In particular, for the $\sim 1460$ ($\sim 260$) SNR system we find that the $1\sigma$ errors on source-frame component masses and redshift agree within $10\,\%$ ($25\,\%$) or better. The largest differences are found in estimates of the redshift, and the lowest in estimates of the primary mass.

For high-redshift IMBH systems with SNR $\sim 30$, we further compared our Fisher estimates with the full PE results of Figs. 9 and 10 of Ref.~\cite{Fairhurst:2023beb}. We find that the $1\sigma$ errors on component masses and redshift are in agreement at the level of $50\,\%$ or better with all the cases considered in those figures. All these comparisons, along with the full PE data of the two IMBH binaries mentioned above, can be found at Ref.~\cite{datarelease}.

Although the Fisher approximation is expected to break down at low SNR or if the actual posterior is multimodal (see e.g. Fig.~8 of Ref.~\cite{Fairhurst:2023beb}), these comparisons show that our estimates of the uncertainties can be trusted at the order-of-magnitude level in most of the parameter space considered in this study.

\bibliography{IMBH_distinguishability_v2}

\end{document}